\documentclass[aip,jcp,twocolumn,superscriptaddress]{revtex4}%
\usepackage{epsfig,amssymb,amsmath,amsthm,amsfonts,amsbsy,mathrsfs}
\usepackage{graphicx,color}
\usepackage{amsmath}
\usepackage{amssymb}%
\usepackage{color}

\usepackage[version=3]{mhchem} % Formula subscripts using \ce{}

\newcommand{\Or}{\mathcal{O}}
\newcommand{\jump}[1]{\big[\hspace{-0.7mm} \big[ #1 \big]
  \hspace{-0.7mm} \big]}
\newcommand{\mean}[1] {\big\{ \hspace{-0.7mm} \big\{ #1 \big\}
  \hspace{-0.7mm} \big\}}
\newcommand{\abs}[1]{\left\lvert#1\right\rvert}

\newcommand{\average}[1]{\left\langle#1\right\rangle}

\newcommand{\mc}[1]{\mathcal{#1}}
\newcommand{\DG}{\mathrm{DG}}
\newcommand{\eff}{\mathrm{eff}}
\newcommand{\ud}{\,\mathrm{d}}
\newcommand{\xc}{\mathrm{xc}}

\begin{document}
\title{DGDFT: A Massively Parallel Method for Large Scale Density Functional Theory Calculations}

\author{Wei Hu}
\thanks{Corresponding author. E-mail: whu@lbl.gov}
\affiliation{Computational Research Division, Lawrence Berkeley
National Laboratory, Berkeley, CA 94720, USA}

\author{Lin Lin}
\thanks{Corresponding author. E-mail: linlin@math.berkeley.edu}
\affiliation{Department of Mathematics, University of California,
Berkeley, CA 94720, USA} \affiliation{Computational Research
Division, Lawrence Berkeley National Laboratory, Berkeley, CA 94720,
USA}

\author{Chao Yang}
\thanks{Corresponding author. E-mail: cyang@lbl.gov}
\affiliation{Computational Research Division, Lawrence Berkeley
National Laboratory, Berkeley, CA 94720, USA}

\date{\today}

\pacs{ }

\begin{abstract}

We describe a massively parallel implementation of the recently
developed discontinuous Galerkin density functional theory (DGDFT)
method, for efficient large-scale Kohn-Sham DFT based electronic
structure calculations. The DGDFT method uses adaptive local basis
(ALB) functions generated on-the-fly during the self-consistent
field (SCF) iteration to represent the solution to the Kohn-Sham
equations. The use of the ALB set provides a systematic way to
improve the accuracy of the approximation. It minimizes the number
of degrees of freedom required to represent the solution to the
Kohn-Sham problem for a desired level of accuracy. In particular,
DGDFT can reach the planewave accuracy with far fewer numbers of
degrees of freedom. By using the pole expansion and selected
inversion (PEXSI) technique to compute electron density, energy and
atomic forces, we can make the computational complexity of DGDFT
scale at most quadratically with respect to the number of electrons
for both insulating and metallic systems. We show that DGDFT can
achieve 80\% parallel efficiency on 128,000 high performance
computing cores when it is used to study the electronic structure of
two-dimensional (2D) phosphorene systems with 3,500-14,000 atoms.
This high parallel efficiency results from a two-level
parallelization scheme that we will describe in detail.

\end{abstract}

\maketitle

\section{Introduction}

Kohn-Sham density functional theory (DFT)\cite{PR_136_B864_1964_DFT,
PR_140_A1133_1965_DFT} is the most widely used methodology for
performing ab initio electronic structure calculations to study the
structural and electronic properties of molecules, solids and
nanomaterials. However, until recently, routine DFT calculations are
only limited to small systems because they have a relatively high
complexity ($\mathcal{O}(N^{2-3})$) with respect to the system size
$N$. As the system size increases, the cost of traditional DFT
calculations becomes prohibitively expensive. Therefore, it is still
challenging to routinely use DFT calculations to treat large-scale
systems that may contain thousands to tens of thousands of atoms.
Although various linear scaling $\mathcal{O}(N)$
methods\cite{RMP_71_1085_1999_ON, IRPC_29_665_2010_ON,
RPP_75_036503_2010_ON} have been proposed for improving the
efficiency of DFT calculations, they rely on the nearsightedness
principle, which leads to exponentially localized density matrices
in real-space for systems with a finite energy gap or at finite
temperature. Furthermore, most of the existing linear scaling DFT
codes, such as SIESTA,\cite{JPCM_14_2745_2002_SIESTA}
OPENMX,\cite{PRB_72_045121_2005_OPENMX}
HONPAS,\cite{IJAC_115_647_2015_HONPAS}
CP2K\cite{CPC_167_103_2005_CP2K} and
CONQUEST,\cite{CPC_177_14_2007_CONQUEST} are based on the contracted
and localized basis sets in the real-space, such as Gaussian-type
orbitals or numerical atomic orbitals.\cite{IRPC_29_665_2010_ON} The
accuracy of methods based on such contracted basis functions are
relatively difficult to improve systematically compared to
conventional uniform basis sets, for example, plane wave basis
set,\cite{PRB_54_11169_1996_PlaneWave} while the disadvantage of
using uniform basis sets is the relatively large number of basis
functions per atom.

Another practical challenge of DFT calculations is related to the
implementation of various approaches to take full advantage of the
massive parallelism available on modern high performance computing
(HPC) architectures. Conventional DFT codes, especially, plane wave
codes, such as VASP,\cite{PRB_47_558_1993_VASP} QUANTUM
ESPRESSO\cite{JPCM_21_395502_2009_QE} and
ABINIT,\cite{CPC_180_2580_2009_ABINIT} have relatively limited
parallel scalability even for large-scale systems involving
thousands of atoms. The strong scaling performance  (i.e., the
change of computational speed with respect to the number of
processing units for a problem of fixed size) of these planewave
based codes is often poor when the number of computational cores
exceeds a few thousands. Nonetheless, improvements have been made in
the past decade. For instance, Qbox\cite{IBM_52_1_2008_Qbox}
demonstrated high scalability to over 1,000 atoms using 131,072
cores on the IBM Blue Gene/L architecture (of which a factor of $8$
is due to $k$-point parallelization). On the other hand, several DFT
software packages based on contracted basis functions have achieved
high parallel performance using linear scaling techniques for
insulating systems. Examples include CP2K\cite{JCTC_8_3565_2012} and
CONQUEST,\cite{JPCM_22_074207_2010} in which linear scaling is
achieved based on parallel sparse matrix
multiplication.\cite{JCC_24_618_2003, JCC_32_1411_2011,
ParallelComput_40_47_2014_BCSR} CP2K has demonstrated calculations
on 96,000 water molecular scaling to 46,656
cores.\cite{JCTC_8_3565_2012} CONQUEST has demonstrated scaling to
over 2,000,000 atoms on 4,096 cores.\cite{JPCM_22_074207_2010}

The recently developed discontinuous Galerkin density functional
theory (DGDFT)\cite{JCP_231_2140_2012_DGDFT} aims at reducing the
number of basis functions per atom while maintaining accuracy
comparable to that of the planewave basis set. This is achieved by
using a set of adaptive local basis (ALB) functions, which are
generated on-the-fly during the self-consistent field (SCF)
iteration. The use of adaptively generated basis functions is also
explored in other software packages such as
ONETEP\cite{SkylarisHaynesMostofiEtAl2005} and recently
BigDFT.\cite{MohrRatcliffBoulangerEtAl2014} One unique feature of
the ALB set is that each ALB function is strictly localized in a
certain element in the real space, and is discontinuous from the
point of view of the global domain. The continuous Kohn-Sham
orbitals and density are assembled from the discontinuous basis
functions using the discontinuous Galerkin (DG)
method.\cite{SIAM_19_742_1982_DG, ArnoldBrezziCockburnEtAl2002} The
ALB set takes both atomic and chemical environmental effects into
account, and can be systematically improved just by increasing one
number (number of ALBs per element). Numerical results suggest that
it can achieve the same level of accuracy obtained by conventional
plane wave calculations with much fewer number of basis functions.
The solution produced by DGDFT is also fully consistent with the
solution of standard Kohn-Sham equations in the limit of a complete
basis set, and the error can be measured by a posteriori error
estimators.\cite{KayeLinYang2015}

The strict spatial locality guarantees that the ALBs can form an
orthonormal basis set and the overlap matrix is therefore an
identity matrix, which requires only the solution of a standard
eigenvalue problem rather than a generalized eigenvalue problem. The
discrete Kohn-Sham Hamiltonian matrix is sparse. Furthermore, the
sparsity pattern bears a resemblance to a block version of finite
difference stencils, and facilitates parallel implementation. The
sparse discrete Hamiltonian matrix allows DGDFT to take advantage of
the pole expansion and selected inversion (PEXSI)
technique.\cite{LinLuYingEtAl2009, JPCM_25_295501_2013_PEXSI,
JPCM_26_305503_2014_PEXSI} The PEXSI method overcomes the $\Or(N^3)$
scaling limit for solving Kohn-Sham DFT, and scales at most as
$\Or(N^2)$ even for metallic systems at room temperature. In
particular, the computational complexity of the PEXSI method is only
$\Or(N)$ for quasi 1D systems, and is $\Or(N^{1.5})$ for quasi 2D
systems. This also makes the DGDFT methodology particularly suitable
for analyzing low-dimensional (quasi 1D and 2D) systems regardless
whether the system is a metal, a semiconductor or an
insulator,\cite{JPCM_25_295501_2013_PEXSI} different from the
near-sightedness assumption in standard linear scaling DFT
calculations.

In this paper, we describe a massively parallel DGDFT method, based
on a two-level parallelization strategy. This strategy results
directly from the domain decomposition nature of the DGDFT method in
which the global computational domain is partitioned into a number
of subdomains from which the ALBs are generated. We demonstrate the
accuracy and high parallel efficiency of DGDFT by using
two-dimensional (2D) phosphorene monolayers with 3,500-14,000 atoms
as examples to show that DGDFT can take full advantage of up to
128,000 cores on a high performance computer.

The rest of paper is organized as follows. After brief introduction
of the DGDFT method, we describe several important implementation
considerations of the massively parallel DGDFT method in
section~\ref{sec:dgdft}. We demonstrate the numerical performance of
DGDFT for 2D phosphorene monolayers in section~\ref{sec:results},
followed by the conclusion in section~\ref{sec:conclusion}.

\section{DGDFT Methodology and Parallel Scheme} \label{sec:dgdft}

%In this section, we briefly present the mathematical foundation and
%algorithmic ingredients of the DGDFT methodology. DGDFT constructs
%adaptive local basis set (ALB) in the discontinuous Galerkin (DG)
%framework~\cite{JCP_231_2140_2012_DGDFT}. We mainly explain why the
%implementation of DGDFT is highly scalable on massively parallel
%computers. Because the sparse Hamiltonian constructed by DGDFT can
%take full advantage of the recently developed pole expansion and
%selected inversion (PEXSI) method\cite{LinLuYingEtAl2009,
%JPCM_25_295501_2013_PEXSI, JPCM_26_305503_2014_PEXSI} to overcome
%the $\Or(N^3)$ scaling of diagonalization methods, it can be used to
%study the electronic structures and ab initio molecular dynamics of
%large-scale systems containing tens of thousand
%atoms.\cite{JCP_141_214704_2014_GNFs, PCCP_2015_ACPNRs}

\subsection{Adaptive local basis set in a discontinuous Galerkin
framework}\label{sec:dgintro}

Since the main focus of this paper is to describe a highly efficient
parallel implementation of the DGDFT methodology, we will not
provide detailed theoretical derivation of the DGDFT method that can
be found in Ref.~\cite{JCP_231_2140_2012_DGDFT}. We will briefly
introduce the basic work flow of DGDFT, and the main steps and major
computational bottlenecks that require parallelization. For detailed
explanation of technical details, we refer readers to
Ref.~\cite{JCP_231_2140_2012_DGDFT} and a forthcoming
publication\cite{DGDFT_MD_2015} concerning the total energy and the
force calculation in DGDFT. Throughout the paper we consider
$\Gamma$-point calculation only. Therefore, the Kohn-Sham orbitals
are real. We assume the system is closed shell. Spin degeneracy is
omitted to simplify the notation in this section, but is correctly
accounted for in the DGDFT code and the numerical simulation.

The goal of the DGDFT method is to find the minimizer of the
Kohn-Sham energy functional, which is a nonlinear functional in
terms of $N$ single particle wavefunctions (orbitals) $\psi_i$. The
minimizer of this functional is sought in a self-consistent field
(SCF) iteration which produces electron density $\rho$ that
satisfies a fixed point (or self-consistent) nonlinear map. In a
standard SCF iteration, a linear eigenvalue problem needs to be
solved in each SCF cycle. As discussed in
Ref.~\cite{JCP_231_2140_2012_DGDFT}, the solution of this linear
eigenvalue problem is usually the computational bottleneck for
solving Kohn-Sham DFT.

\begin{figure}[htbp]
\begin{center}
\includegraphics[width=0.5\textwidth]{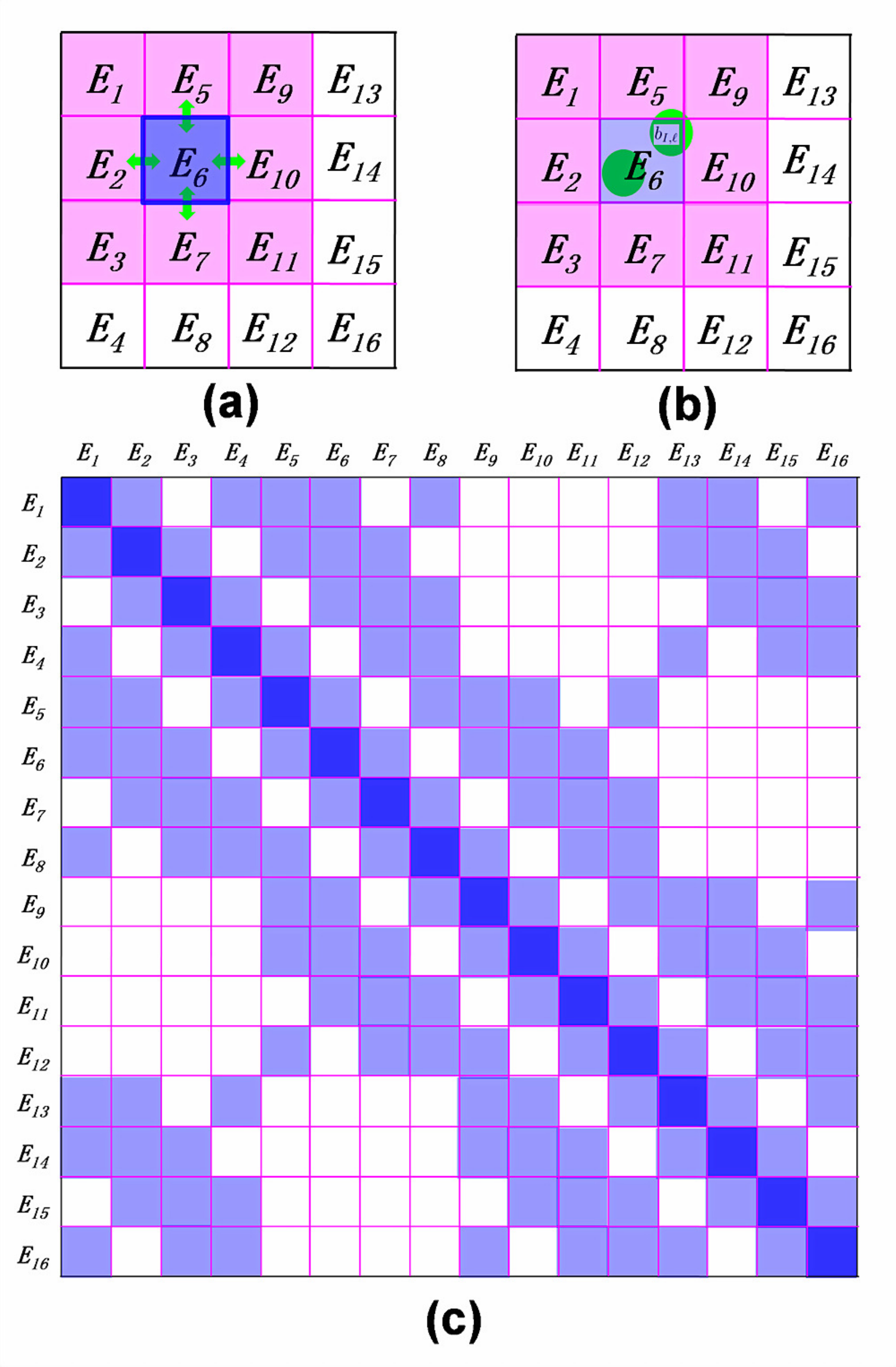}
\end{center}
\caption{(Color online) A model system in 2D partitioned into $16$
(4 $\times$ 4) equal sized elements.  (a) An extended element $Q_6$
associated with the element $E_{6}$, and $Q_{6}$ includes elements
$E_1$, $E_2$, $E_3$, $E_5$, $E_6$, $E_7$ $E_9$, $E_{10}$ and
$E_{11}$. The four surfaces of $E_{6}$ on which boundary integrals
are computed are highlighted by arrows. (b) Two possible cases of
projection vectors of the nonlocal pseudopotential, one contained a
element, and another contained in multiple (here $4$) elements. (c)
The block structure of DG Hamiltonian matrix with blocks with
nonzero values highlighted.} \label{fig:HDG2D} \end{figure}

In the DGDFT method, we partition the global computational domain
$\Omega$ into a number of subdomains (called elements), denoted by
$E_{1},\ldots,E_{M}$. An example of partitioning the global domain
of a model problem into a number of 2D elements is given in
Fig.~\ref{fig:HDG2D}. We use $\mc{T}$ to denote the collection of
all elements. In the current version of DGDFT, we use periodic
boundary conditions to treat both molecule and solids. However, the
method can be relatively easily generalized to other boundary
conditions such as Dirichlet boundary condition. Therefore each
surface of the element must be shared between two neighboring
elements, and $\mc{S}$ denotes the collection of all the surfaces.

Each Kohn-Sham orbital is expanded into a linear
combination of adaptive local basis functions (ALBs) $\varphi_{K,j}$, i.e.,
\begin{equation}\label{eqn:psiexpand}
  \psi_i(x) = \sum_{K=1}^{M} \sum_{j=1}^{J_K} c_{i; K, j}
  \varphi_{K, j}(x),
\end{equation}
where $\varphi_{K, j}(x)$ is the $j$-th ALB in the element $E_{K}$
that has nonzero support only in $E_{K}$, and $J_{K}$ is the total
number of ALBs in $E_{K}$. The basis function $\varphi_{K,j}$ is not
necessarily continuous across the boundary of $E_K$. Because each
$\varphi_{K,j}$ is assumed to be square integrable within $E_K$ and
is zero outside of $E_K$, its inner product with other function can
be defined in terms of an $L^{2}$ inner product defined on $E_K$. As
a result, a natural global inner product between quantities such as
$\psi_i$, which we denote by $\average{\cdot, \cdot}_{\mc{T}}$, can
be taken as the sum of local $L^{2}$ inner products defined for all
elements. Similarly, we can also define a global surface inner
product, denoted by $\average{\cdot, \cdot}_{\mc{S}}$, as the sum of
local $L^{2}$ surface inner products defined on all surfaces of all
elements. The ALB set is orthonormal, i.e., for
$K,K'=1,\ldots,M,\quad j=1,\ldots,J_{K},\quad j'=1,\ldots,J_{K'}$,
\begin{equation}
  \average{\varphi_{K, j}, \varphi_{K', j'}}_{\mc{T}} =
\delta_{K,K'}\delta_{j,j'}.
  \label{eqn:orthonormal}
\end{equation}

Because the expansion of $\psi_{i}$ contains at least two ALBs
localized in two adjacent elements with a shared surface on which
neither ALB is continuous, the values of $\psi_{i}$ on two sides of
the surface can be different. This difference calls for the notion
of average of gradient $\mean{\nabla\psi_i}$ and the concept of jump
of function value $\jump{\psi_i}$ defined on the surface. The use of
average and jump operators distinguishes DGDFT from other KS-DFT
solvers.

Using these notations, the total energy functional to be minimized
with respect to the $N$ Kohn-Sham orbitals $\{\psi_{i}\}$
subject to orthonormality condition can be written as
\begin{equation}\label{eqn:DGvar}
  \begin{split}
    E_{\DG}(\{\psi_i\}) = &\frac{1}{2} \sum_{i=1}^N \average{\nabla
    \psi_i , \nabla \psi_i}_{\mc{T}}
    + \average{ V_{\eff}, \rho }_{\mc{T}}  \\
    &+ \sum_{I=1}^{N_A}\sum_{\ell=1}^{L_{I}} \gamma_{I,\ell} \sum_{i=1}^N
    \abs{\average{b_{I,\ell}(\cdot-R_{I}), \psi_i}_{\mc{T}}}^2 \\
    &- \sum_{i=1}^N
    \average{\mean{\nabla\psi_i}, \jump{\psi_i}}_{\mc{S}}\\
    &+ \alpha \sum_{i=1}^N \average{\jump{\psi_i},
    \jump{\psi_i}}_{\mc{S}},
  \end{split}
\end{equation}
where we use $V_{\eff}$ to denote the effective one-body potential
(including local pseudopotential $V_{\text{loc}}$, Hartree potential
$V_{H}$ and the exchange-correlation potential $V_{\xc}[\rho]$), the
terms that contain $b_{I,\ell}$ and $\gamma_{I,\ell}$ correspond to
the nonlocal pseudopotential in the Kleinman-Bylander
form,\cite{PRL_48_1425_1982} and $\alpha$ is an adjustable penalty
parameter to ensure that Eq.~\eqref{eqn:DGvar} has a well defined
ground state energy. For each atom $I$, there are $L_I$ functions
$\{b_{I,\ell}\}$, called projection vectors of the nonlocal
pseudopotential. The parameters $\{\gamma_{I,\ell}\}$ are real
valued scalars. We refer the readers to
Ref.~\cite{JCP_231_2140_2012_DGDFT} for more detailed explanation of
the notation and theory.

The procedure for constructing the ALBs and a detailed example of
the ALBs will be given later in this subsection. For now we just treat
Eq.~\eqref{eqn:psiexpand} as an ansatz for representing the Kohn-Sham
orbitals. Minimizing the coefficients $\{c_{i;K,j}\}$ subject to
orthonormality condition gives rise to the Euler-Lagrange equation of
Eq.~\eqref{eqn:DGvar} which is a linear eigenvalue problem
\begin{equation}
  \sum_{K,j} H^{\DG}_{K', j';  K, j} c_{i; K,j} = \lambda_{i} c_{i;K',j'}.
  \label{eqn:ELeq}
\end{equation}
Here $H^{DG}$ is the discrete Kohn-Sham Hamiltonian matrix, with
matrix entries given by
\bgroup
%\small
%\begin{widetext}
\begin{equation}\label{eqn:DGHamiltonian}
  \begin{split}
    & H^{\DG}_{K', j';  K, j}\\
    ={}&\Bigl(\frac{1}{2} \average{\nabla
    \varphi_{K, j'}, \nabla \varphi_{K, j}}_{\mc{T}}
    + \average{\varphi_{K, j'}, V_{\eff}\varphi_{K, j}}_{\mc{T}}
    \Bigr) \delta_{K,K'}\\
    &+ \Bigl(\sum_{I,\ell} \gamma_{I,\ell} \average{\varphi_{K', j'},
    b_{I,\ell}}_{\mc{T}} \average{b_{I,\ell}, \varphi_{K, j}}_{\mc{T}}
    \Bigr) \\
    &+ \Bigl(- \frac{1}{2} \average{\jump{\varphi_{K', j'}},
    \mean{\nabla\varphi_{K, j}}}_{\mc{S}} \\
    &\phantom{+ \Bigl(} - \frac{1}{2} \average{\mean{\nabla\varphi_{K', j'}}, \jump{\varphi_{K,
    j}}}_{\mc{S}}\\
    &\phantom{+ \Bigl(} + \alpha \average{\jump{\varphi_{K', j'}}, \jump{\varphi_{K,
    j}}}_{\mc{S}}\Bigr).
  \end{split}
\end{equation}
%\end{widetext}
\egroup The $H^{\DG}$ matrix can be naturally partitioned into
matrix blocks as sketched in Fig.~\ref{fig:HDG2D}. We call the
submatrix $H^{\DG}_{K',:;K,:}$ of size $J_{K'}\times J_{K}$ the
$(K',K)$-th matrix block of $H^{\DG}$, or $H^{\DG}_{K';K}$ for
short. The terms are grouped by three parentheses on the right hand
side of Eq.~\eqref{eqn:DGHamiltonian} to reflect different
contributions to the DG Hamiltonian matrix, and will be treated
differently in our parallel implementation of the method. The first
group originates from the kinetic energy and the local
pseudopotential, and only contributes to the diagonal blocks
$H^{\DG}_{K,K}$. The second group comes from nonlocal
pseudopotentials, and contributes to both the diagonal and
off-diagonal blocks of $H^{\DG}$. Since a projection vector of the
nonlocal pseudopotential is spatially localized, we require the
dimension of every element along each direction (usually on the
order of $6\sim 8$ Bohr) to be larger than the size of the nonzero
support of each projection vector (usually on the order of $2\sim 4$
Bohr). Thus, the nonzero support of each projection vector can
overlap with at most $2^{d}$ elements as shown in
Fig.~\ref{fig:HDG2D} (b) ($d=1,2,3$). As a result, each nonlocal
pseudopotential term may contribute both to the diagonal and the
off-diagonal blocks of $H^{\DG}$. The third group consists of
contributions from boundary integrals, and can also contribute to
both the diagonal and off-diagonal blocks of $H^{\DG}$. Each
boundary term involves only two neighboring elements by definition
as plotted in Fig.~\ref{fig:HDG2D} (a). In summary, $H^{\DG}$ is a
sparse matrix and the nonzero matrix blocks correspond to
interactions between neighboring elements (Fig.~\ref{fig:HDG2D}).

After constructing the $H^{\DG}$ matrix and solving the eigenvalue
problem~\eqref{eqn:ELeq}, the electron density required to evalute the
effective potential $V_{\eff}$ in Eq.~\eqref{eqn:DGvar} can be obtained from
\begin{equation}
  \label{eqn:density}
  \rho(x) = \sum_{i=1}^N \abs{\psi_{i}(x)}^2.
\end{equation}

It is well known that \eqref{eqn:density} is not the only way to
compute the electron density. An alternative approach which does not
require computing eigenvalues and eigenvectors of $H_{DG}$ is the
pole expansion and selected inversion (PEXSI) method
\cite{LinLuYingEtAl2009, JPCM_25_295501_2013_PEXSI,
JPCM_26_305503_2014_PEXSI}. To make use of the PEXSI method, we need
to express $\rho(x)$ in terms of selected blocks of the density
matrix represented in the ALB set (or density matrix for short in
the following discussion). To be more precise, this density matrix
has the form
\begin{equation}
  P_{K,j;K',j'} = \sum_{i=1}^{N} c_{i;K,j} c_{i;K',j'},
  \label{eqn:densitymatrix}
\end{equation}
and can be accurately approximated as a matrix function of $H_{DG}$
without knowing $c_{i;K,j}$ explicitly.

Using the density matrix, we can express the electron density as
\begin{equation}
  \label{eqn:newdensity}
  \begin{split}
  &\rho(x)\\
  ={} &\sum_{K=1}^{M} \sum_{j=1}^{J_{K}}
  \sum_{K'=1}^{M} \sum_{j'=1}^{J_{K'}} \varphi_{K,j}(x)
  \varphi_{K',j'}(x) \left(\sum_{i=1}^{N} c_{i; K,j} c_{i; K',
  j'}\right)\\
  ={}& \sum_{K=1}^{M} \sum_{j=1}^{J_{K}} \sum_{j'=1}^{J_{K}}
  \varphi_{K,j}(x) \varphi_{K,j'}(x) P_{K,j;K,j'}.
  \end{split}
\end{equation}
Here we have used the fact that each function $\varphi_{K,j}(x)$ is
strictly localized in the element $E_{K}$ to eliminate the cross
terms involving both $K$ and $K'$. As a result, the selected blocks,
or more specifically, the diagonal blocks of the density matrix
$P_{K;K}\equiv P_{K,:;K,:}$ are needed to evaluate the electron
density. This is a key feature of this expression that makes it
possible to use the PEXSI method as will be discussed in
section~\ref{sec:pexsi}. After self-consistency of the electron
density is achieved in the SCF iteration, the total energy and
atomic forces can also be evaluated using the PEXSI method.

We have demonstrated that high accuracy in the total energy and
atomic forces can be achieved with a very small number (4$-$40) of
basis functions per atom in DGDFT, compared to fully converged
planewave calculations.\cite{JCP_231_2140_2012_DGDFT, DGDFT_MD_2015}
Besides selected blocks of the density matrix (DM), the Helmholtz
free energy and the atomic force can be evaluated by computing
selected blocks of the free energy density matrix (FDM) and the
energy density matrix (EDM), respectively. The technique for
computing EDM and FDM is very similar to that for computing the
DM.\cite{JPCM_25_295501_2013_PEXSI}

One notable feature of the ALB set is that they are generated on the
fly, and is adaptive not only to the atomic but also the
environmental information. In order to construct ALBs, we introduce,
for each element $E_K$, an extended element $Q_K$ that contains
$E_K$ and a buffer region surrounding $E_K$. We define
$V_{\eff}^{Q_{K}}=V_{\eff}\vert_{Q_{K}}$ to be the restriction of
the effective potential at the current SCF step to $Q_{K}$, and
$V_{\text{nl}}^{Q_{K}}=V_{\text{nl}}\vert_{Q_{K}}$ to be the
restriction of the nonlocal potential to $Q_{K}$. These restricted
potentials define a local Kohn-Sham linear eigenvalue problem on
each extended element $Q_K$:
\begin{equation}\label{eqn:localproblem}
    \begin{split}
      &\widetilde{H}_{\eff}^{Q_{K}} \widetilde{\varphi}_{K,j} = \left(-\frac12 \Delta + V_{\eff}^{Q_{K}}
   +V_{\text{nl}}^{Q_{K}}\right) \widetilde{\varphi}_{K,j} = \lambda_{K,j}
        \widetilde{\varphi}_{K,j},\\
        &\int_{Q_{K}}
        \widetilde{\varphi}_{K,j}(x)\widetilde{\varphi}_{K,j'}(x) \ud x =
        \delta_{jj'}.
    \end{split}
\end{equation}
This linear eigenvalue problem can be discretized by using
traditional basis sets such as plane waves and solved by an
iterative method such as the locally optimal block preconditioned
conjugate gradient (LOBPCG) method. We remark that SCF iterations
need not and should not be performed within each element $Q_{K}$.
The lowest $J_{K}$ eigenvalues $\{\lambda_{K,j}\}_{j=1}^{J_{K}}$ and
the corresponding eigenfunctions
$\{\widetilde{\varphi}_{K,j}\}_{j=1}^{J_{K}}$ are computed on
$Q_{K}$. We then restrict $\{\widetilde{\varphi}_{K,
j}\}_{j=1}^{J_{K}}$ from $Q_{K}$ to $E_K$. The truncated vectors are
not necessarily orthonormal. Therefore, we orthonormalize the set of
truncated eigenvectors to obtain $\{\varphi_{K,j}\}_{j=1}^{J_K}$. We
then set each $\varphi_{K,j}$ to zero outside of $E_K$, so that it
defined over the entire domain, but is in general discontinuous
across the boundary of $E_K$. These functions constitute the ALB set
that we use to represent the Kohn-Sham Hamiltonian.  Because they
satisfy the orthonormality condition~\eqref{eqn:orthonormal}, the
overlap matrix corresponding to the ALB set is an identity matrix.
Hence, no generalized eigenvalue problem with a potentially
ill-conditioned overlap matrix need to be solved. For more details
of the construction of ALBs, such as the restriction of the
potential and the choice of boundary conditions for the local
eigenvalue problem~\eqref{eqn:localproblem}, we refer readers to
Ref.~\cite{JCP_231_2140_2012_DGDFT}.

\begin{figure}[htbp]
\begin{center}
\includegraphics[width=0.5\textwidth]{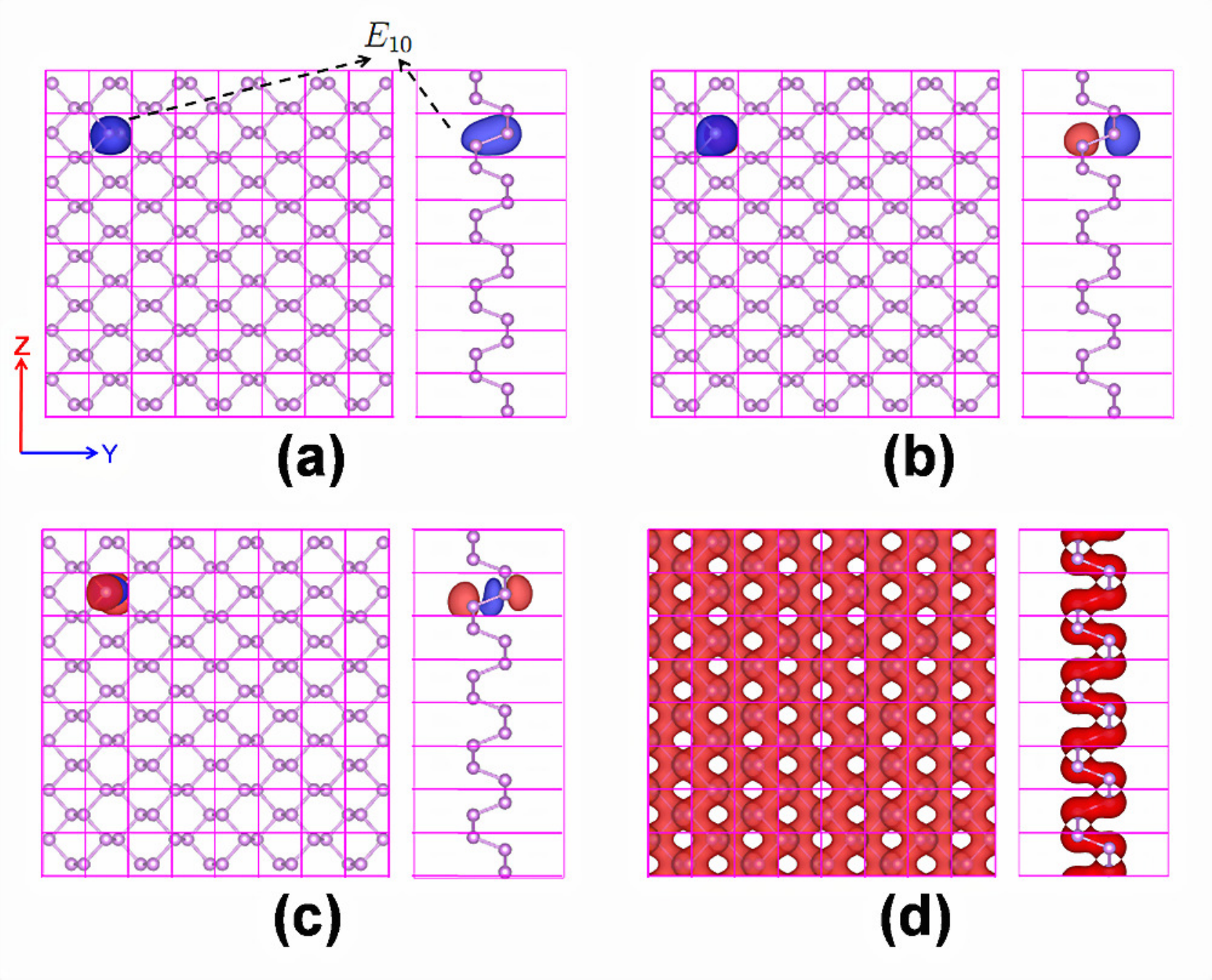}
\end{center}
\caption{(Color online) The isosurfaces (0.04 Hartree/Bohr$^3$) of
the first three ALB functions belonging to the tenth element
($E_{10}$), (a) $\phi$$_1$, (b) $\phi$$_2$, (c) $\phi$$_3$, and (d)
the electron density $\rho$ across in top and side views in the
global domain in the example of P$_{140}$. The red and blue regions
indicate positive and negative isosurfaces, respectively. There are
64 elements and 80 ALB functions in each element in the P$_{140}$
system.} \label{fig:Structure}
\end{figure}

As an example, Fig.~\ref{fig:Structure} shows the ALBs of a 2D
phosphorene monolayer with $140$ phosphorus atoms (P$_{140}$). This
is a two-dimensional system, and the global computational domain is
partitioned into $64$ equal sized elements along the Y and Z
directions, respectively. For instance, the extended element
$Q_{10}$ associated with the element $E_{10}$ contains elements
$E_{1}$, $E_{2}$, $E_{3}$, $E_{9}$, $E_{11}$, $E_{17}$, $E_{18}$ and
$E_{19}$. We show the isosurfaces of the first three ALB functions
for this element in Fig.~\ref{fig:Structure}(a)-(c). Each ALB
function shown is strictly localized inside $E_{10}$ and is
therefore discontinuous across the boundary of elements. On the
other hand, each ALB function is delocalized across a few atoms
inside the element since they are obtained from eigenfunctions of
local Kohn-Sham Hamiltonian. Although the basis functions are
discontinuous, the electron density is well-defined and is very
close to be a continuous function in the global domain
(Fig.~\ref{fig:Structure}(d)).

\begin{figure}[htbp]
\begin{center}
\includegraphics[width=0.5\textwidth]{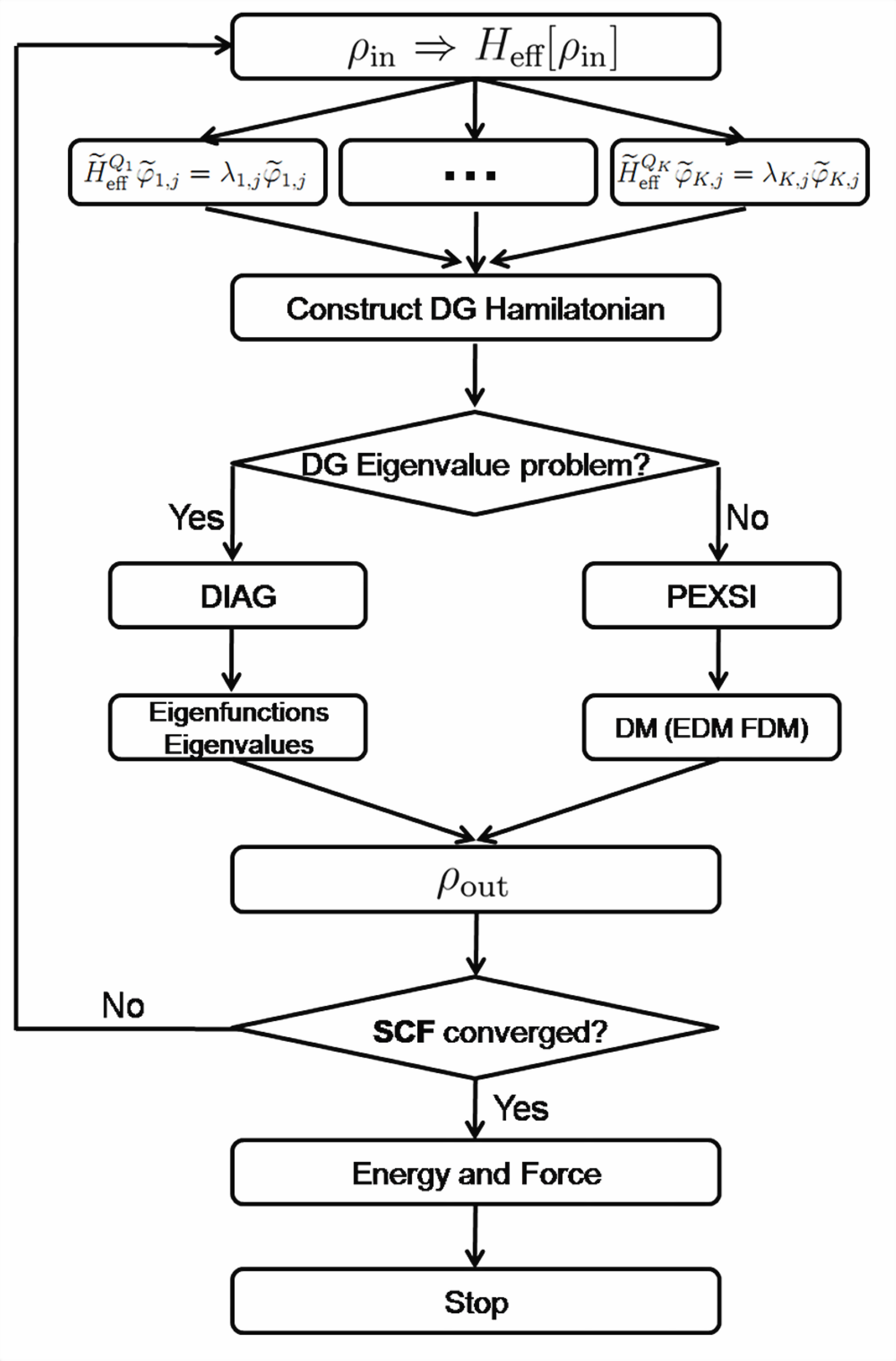}
\end{center}
\caption{(Color online) Flowchart of the DGDFT method. There are three
time consuming steps: the generation of ALB functions, the
construction of DG Hamiltonian matrix,  and the
evaluation of the approximate charge density by either diagonalizing
the DG Hamiltonian (DIAG) or by using the PEXSI method. $H_{\eff}$
and $\widetilde{H}_{\eff}^{Q_{K}}$ represent the global Kohn-Sham
Hamiltonian operator for a given electron density, and the local
Kohn-Sham Hamiltonian on the extended element $Q_{K}$, respectively.}
\label{fig:Flowchart}
\end{figure}

To summarize, the flowchart of the DGDFT method for solving
Kohn-Sham DFT is shown in Fig.~\ref{fig:Flowchart}.

\subsection{Two level parallelization strategy}\label{sec:parallel}

%The DGDFT framework naturally allows two levels of parallelization
%on elements and adopt 2D MPI (message passing interface)
%communications (N$_{\text{DGRow}}$ * N$_{\text{DGCol}}$)
%respectively for inter-element and intra-element parallelization as
%shown in Figure~\ref{fig:DGDFT1}. For each (or extended) element in
%subdomain, the computation of eigenfunctions for the local Kohn-Sham
%Hamiltonian can be parallelized similar to how a regular Kohn-Sham
%DFT solver with planewave basis sets is parallelized on
%N$_{\text{DGRow}}$ processors. The ALB functions matrix is in two
%dimensions of \#Grid and \#ALB, which is determined by the kinetic
%energy cutoff ($E_{\text{cut}}^{\text{wfc}}$) of the number of ALB
%functions contained in extended element. The relationship between
%$E_{\text{cut}}^{\text{wfc}}$ and the number of uniform grid points
%along $i$th direction (N$i$), where $i$ $\in$ ${x,y,z}$, can be
%written as N$i$ = $\sqrt{2E_{\text{cut}}^{\text{wfc}}}$L$i$/$\pi$,
%where L$i$ is the dimension of the extended element along the $i$th
%direction.

The DGDFT framework naturally lends itself to a two level
parallelization strategy. At the coarse grained level, we distribute
work among different processors by elements. We call this level of
concurrency the \textit{inter-element} parallelization. Within each
element, the eigenvalue solvers on each local (extended) domain and
the construction of the DG Hamiltonian matrix can be further
parallelized. This level of parallelization is called the
\textit{intra-element} parallelization. We use the Message Passing
Interface (MPI) to handle data communication.
\begin{figure}[htbp]
\begin{center}
\includegraphics[width=0.4\textwidth]{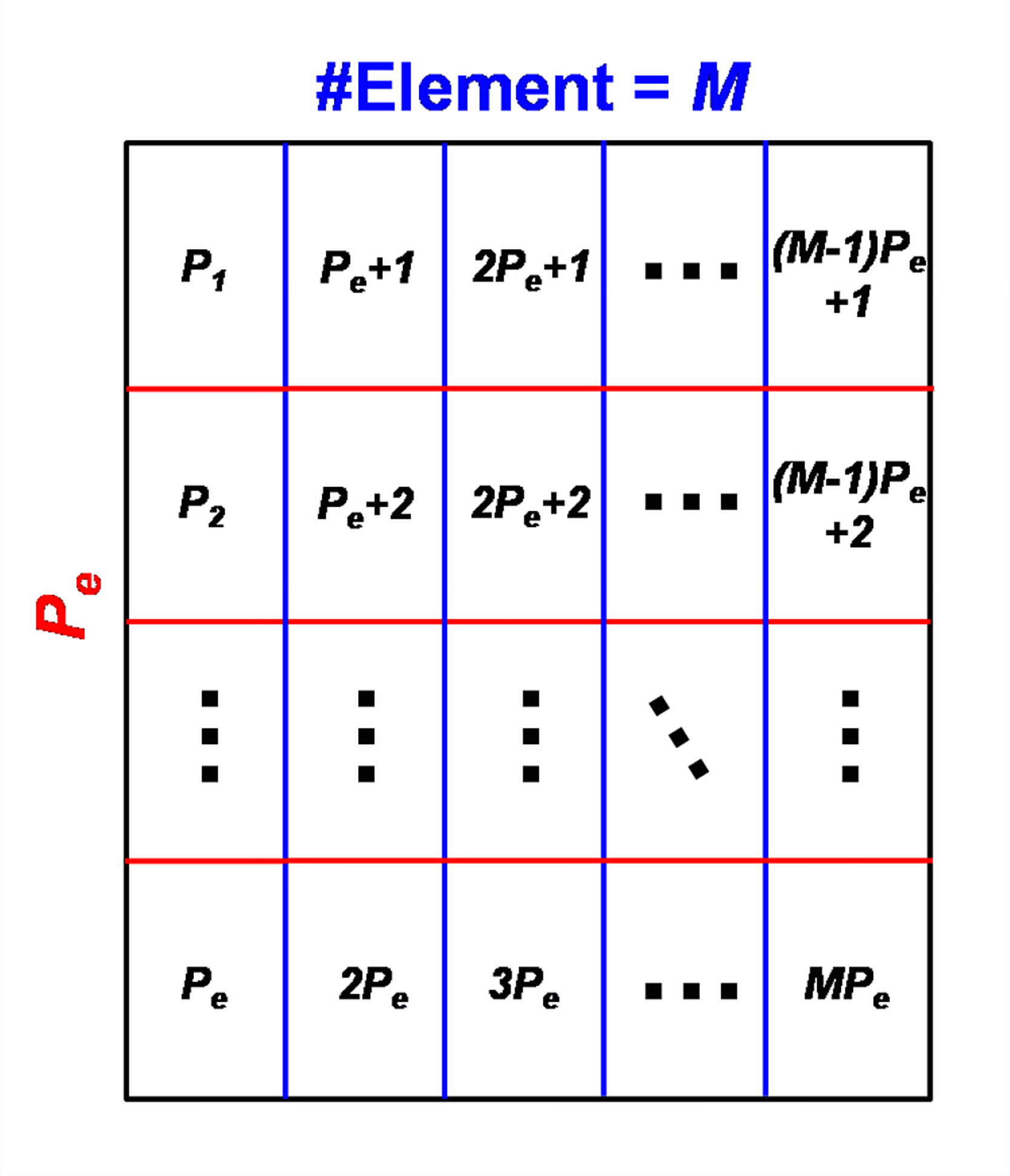}
\end{center}
\caption{(Color online) Global processor grid for implementing the
two level parallelization in the DGDFT method. $M$ is the number of
elements in the systems. $P_{e}$ is the number of processors given
for each element.} \label{fig:DGDFT1}
\end{figure}

We assume that the total number of processors is $P_{\mathrm{tot}} =
M \times P_{e}$, where $M$ is the number of elements, and $P_{e}$ is
the number of processors assigned to each element. We partition
these processors into a 2D logical processor grid following a column
major order as shown in Fig.~\ref{fig:DGDFT1}. We call this 2D
processor grid the global processor grid to distinguish it from
other processor grids employed in various parts of the massively
parallel DGDFT method. Each row (column) communicator of this grid
is called a global row (column) communicator. The processors with
ranks $(K-1)P_{e}+1$ to $K P_{e}$ ($K=1,\ldots, M)$ are in the
$K$-th global column processor group, and are assigned to element
$E_{K}$ for intra-element parallelization. Similarly, the processors
with ranks $i,P_{e}+i,\ldots,(M-1)P_{e}+i$ ($i=1,\ldots,P_{e}$) are
in the $i$-th global row processor group. When a very large number
of processors are used, it is important to avoid global all-to-all
communication as much as possible in order to reduce communication
cost and achieve parallel scalability. Therefore by dividing
processors into column and row processor groups, we can restrict
most of the communication within a global column or row group.

Based on the intra-element parallelization strategy to be detailed
below, the maximum number of processors that can be effectively used
for a single element depends on the number of basis functions to be
generated on a single element, which is usually on the order of
$100$. It can be as large as a few hundreds. The level of
concurrency that can be achieved in the inter-element
parallelization is determined by the number of elements. The maximum
number of processors that can be employed by DGDFT is therefore
determined by the number of basis functions per element multiplied
by the number of elements, which is equal to the dimension of the DG
Hamiltonian matrix. For example, the P$_{140}$ system contains 140
phosphorus atoms and is partitioned into 64 (8 $\times$ 8) elements.
There are about 2 atoms in each element and we use 50 basis per
element. The maximum number of processors that can be effectively
used for this system is 3,600. For the 2D phosphorene test problems
(P$_{3500}$ and P$_{14000}$) used in this work, we partition the
phosphorene systems into 1,600 (40 $\times$ 40) and 6,400 (80
$\times$ 80) elements respectively. The maximum number of processors
that can be effectively used for these systems are 80,000 and
320,000, respectively. Currently, we are often limited by the number
of processors available on the existing high performance computers
such as the Edison system at NERSC. The maximum number of processors
we can use is 128,000.

\textbf{Generation of the ALB functions}

\begin{figure}[htbp]
\begin{center}
\includegraphics[width=0.5\textwidth]{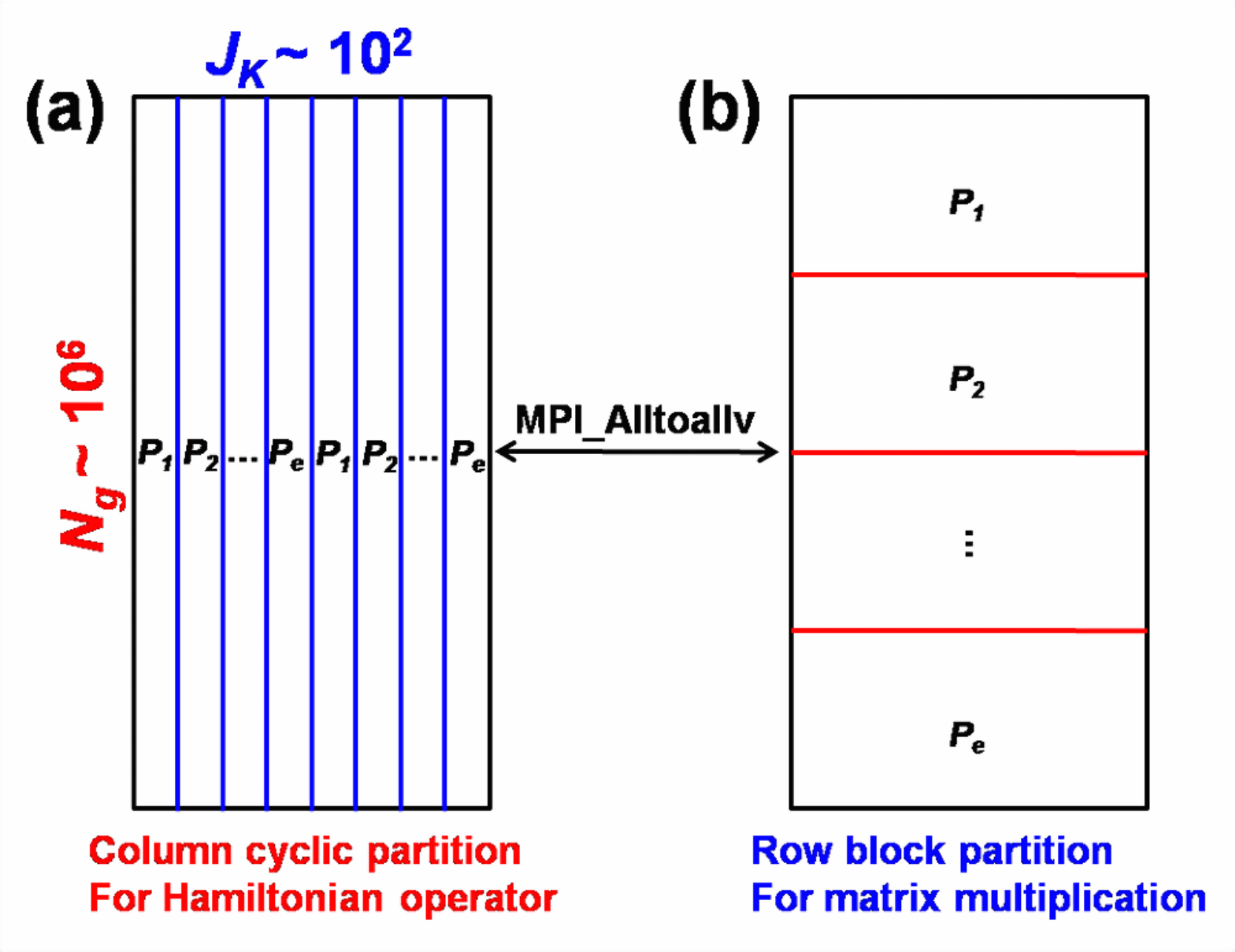}
\end{center}
\caption{(Color online)Two different types of partition of
$\Psi_{K}$. The partition shown in (a) is used for computing
$\widetilde{H}_{\eff}^{Q_{K}}\Psi_{K}$. The partition shown in (b)
is used when matrix-matrix multiplications of the form
$\Phi^T\Psi_{K}$ and $\Psi_{K}C$ are performed.} \label{fig:DGDFT2}
\end{figure}

For each extended element, the computation of eigenfunctions for the
local Kohn-Sham Hamiltonian can be parallelized in a way similar to
the parallelization of a standard planewave based Kohn-Sham DFT solver.
The local Kohn-Sham orbitals form a local dense matrix
$\Psi_{K}$ of dimension $N_{g}\times J_{K}$. Here $J_{K}$ is the
number of ALBs to be computed, and $N_{g}$ is the total number of
grid points required to represent each local Kohn-Sham orbital in
the real space, which is determined from the kinetic energy cutoff
$E_{\text{cut}}^{\text{wfc}}$ by the following rule:
\begin{equation}
(N_{g})_{i} = \frac{\sqrt{2E_{\text{cut}}^{\text{wfc}}} L_{i}}{\pi}.
  \label{eqn:Ngecut}
\end{equation}
Here $L_{i}$ is the length of the extended element along the $i$-th
coordinate direction. The total number of grid points is
$N_{g}=\prod_{i=1}^{3} (N_{g})_{i}$. In a typical calculation
$N_{g}\sim 10^6$ and $J_{K}\sim 10^2$ as shown in
Fig.~\ref{fig:DGDFT2}.

On each extended element, we use the LOBPCG method\cite{Knyazev2001}
to solve the local Kohn-Sham orbitals. The main operations involved
in the LOBPCG solver are: 1) The application of the local
Hamiltonian operator $\widetilde{H}_{\eff}^{Q_{K}}$ to $\Psi_{K}$. 2) Dense
matrix-matrix multiplication of the form $\Phi^{T}\Psi$, where
matrices $\Phi,\Psi$ have the same dimension as the Kohn-Sham
orbitals $\Psi_{K}$. 3) Dense matrix-matrix multiplication of the
form $\Phi C$, where $\Phi$ has the same dimension as $\Psi_{K}$,
and $C$ is a small matrix on the order of $J_{K}\times J_{K}$. 4)
The diagonalization of matrices of size $\mathcal{O}(J_{K})$ using
the Rayleigh-Ritz procedure. 5) The application of a preconditioner
operator.

It should be noted that the these different types of operations
require different data distribution and task parallelization
strategies.  Operation 1) requires applying the Laplacian operator
via the Fast Fourier Transform (FFT), the local pseudopotential
operator and the nonlocal pseudopotential operator to $\Psi_{K}$.
This can be done easily if the orbitals (i.e., columns of
$\Psi_{K}$) are partitioned along the column direction (see
Fig.~\ref{fig:DGDFT2} (a)). Since each processor holds all entries
of an orbital, the FFT can be done in the same way as in a
sequential implementation. We further assume that each processor
associated with the extended element $Q_{K}$ has a copy of the local
pseudopotential $V_{\eff}^{Q_{K}}$, and the nonlocal pseudopotential
$V_{\text{nl}}^{Q_{K}}$. Therefore all operations described in 1)
can be performed in the same way as those in a sequential
implementation.

The most efficient way to parallelize operations 2) and 3) is to
partition $\Psi_{K}$ by row blocks as shown in Fig.~\ref{fig:DGDFT2}
(b). This is because for operation 2), one can compute the matrix
inner product of each block locally on each processor, and then use
MPI\_Allreduce among the $P_{e}$ processors to sum up local products
of size $J_{K}\times J_{K}$.  For operation 3), there is no
communication at all if all $P_{e}$ processors have a local copy of
the $C$ matrix. Partitioning $\Psi_{K}$ by columns would incur more
communication cost and make this part of the computation less
scalable. Since in each LOBPCG iteration we apply the Hamiltonian
operator to $\Psi_{K}$ once, but perform $12$ matrix-matrix
multiplication of operation type 2) and $7$ matrix-matrix
multiplication of operation type 3), it is worthwhile to switch back
and forth between a column partition and row partition of $\Psi_K$
in between the first and other types of operations performed in each
LOBPCG iteration. This can be achieved by using a MPI\_Alltoallv
call.

For operation 4), since $J_{K}$ is usually on the order of hundreds
in practice, we solve the Rayleigh-Ritz problem and perform subspace
diagonalization locally on each processor. Numerical experiments
indicate that this sequential part usually takes around or less
than 1 second.

Finally, in our implementation we use the preconditioner proposed in
Ref.~\cite{PayneTeterAllenEtAl1992} to accelerate the convergence of
LOBPCG. The preconditioner can be easily applied to different
orbitals simultaneously without communication. Thus a column
partition of $\Psi_{K}$ is suitable for applying the preconditioner
in parallel.

Once the Kohn-Sham orbitals $\Psi_{K}$ are constructed through
LOBPCG, they are restricted from the extended element $Q_{K}$ to the
element $E_{K}$. After orthonormalizing columns of the restricted
$\Psi_K$, we obtain the ALBs denoted by
$\Phi_{K}=[\varphi_{K,1},\ldots,\varphi_{K,J_{K}}]$. We remark that
it is not necessary to compute the local wavefunctions $\Psi_{K}$ to
full accuracy before the electron density becomes self consistent in
the SCF iteration. As the accuracy of the electron density improves
during the SCF cycle, a more accurate $\Psi_{K}$ can be obtained
from running a few iterations of the LOBPCG procedure that uses the
$\Psi_K$ returned from the previous SCF iteration as a starting
guess. In practice, we find that using $3$ preconditioned LOBPCG
iterations per SCF iteration is often sufficient to achieve rapid
convergence of the SCF procedure. Our numerical results indicate
that the overall procedure for generating the ALBs is highly
efficient. For instance, for the phosphorene P$_{3500}$ system, the
total time for generating the ALBs is only 1.35 and 0.71 sec by
using 12,800 and 25,600 processors, respectively.

%On each extended element, we use the LOBPCG method to solve local
%eigenvalue problem to generate the ALB functions, including the
%matrix-matrix multiplication, matrix vector multiplication,
%Rayleigh-Ritz procedure and Hamiltonian operator (FFT, local and
%nonlocal pseudopotential). Because we need to solve a eigenvalue
%problem containing only tens of atoms in this subdomain (\#Grid
%$\sim$ 10$^6$ and \#ALB $\sim$ 10$^2$), we implement straightforward
%parallelization over grids for row block partition and
%parallelization over ALB bands for column cyclic partition shown in
%Figure~\ref{fig:DGDFT2}. We parallelize the matrix-vector
%multiplication (involving FFT etc) using column cyclic partition
%since the number of grids on the local domain is still relatively
%small and sequential algorithm is sufficiently efficient for a
%single band. We find that the same column partition is much less
%efficient for GEMM. In fact since the ALB functions for each element
%form a tall skinny matrix, row partition is much more efficient.
%Finally the typical size of the number of ALB functions is 10$^2$,
%so the cost of the projected Rayleigh-Ritz problem is usually small
%and is solved via LAPACK sequentially. Numerical experience is that
%all the sequential work is less than 1 second even for ten of
%thousands atoms system. Between each matrix-vector multiplication,
%the change from row to column partition is done via MPI\_Alltoallv.

\textbf{Construction of the DG Hamiltonian}

Due to the spatial locality of the ALBs, the DG Hamiltonian matrix
in Eq.~\eqref{eqn:DGHamiltonian} is a sparse matrix and has a block
structure that can be naturally distributed among different column
processor groups assigned to different elements as shown in
Fig.~\ref{fig:HDG2D} (c), i.e. the processors assigned to the
element $E_{K}$ assembles the $K$-th block row of $H^{\DG}$. The
construction of the DG Hamiltonian matrix consists of the evaluation
of volume integrals within each element and surface integrals at the
boundary of different elements. To achieve high accuracy, all
integrals in \eqref{eqn:DGHamiltonian} are evaluated by a Gaussian
quadrature defined on a Legendre-Gauss-Labotto (LGL) grid in each
element. The gradients of ALBs sampled on the LGL grid points and
denoted by $\nabla \Phi_K$, can be evaluated by applying
differentiation matrices to $\Phi_K$ along the $x,y,z$ directions,
respectively. To compute $\nabla \Phi_K$ efficiently, both $\Phi_K$
and $\nabla \Phi_K$ should be partitioned and distributed by columns
among processors assigned to $E_K$. We refer readers to e.g.
Ref.~\cite{Trefethen2000} for details of constructing
differentiation matrices associated with Gaussian quadratures.

The construction of the $K$-th block row of $H^{\DG}$ consists of
the following three steps that correspond to the three terms grouped
by parentheses on the right hand side of
Eq.~\eqref{eqn:DGHamiltonian}. 1) Compute contributions to the
diagonal matrix blocks from the kinetic energy and $V_{\eff}$ terms.
2) Compute contributions to the diagonal and off-diagonal matrix
blocks from the nonlocal pseudopotential term. 3) Compute
contributions to the diagonal and off-diagonal matrix blocks from
the boundary integral terms.

In step 1), it is generally more efficient to partition columns of
$\Phi_K$ among different processors as shown in
Fig.~\ref{fig:DGDFT2}(a) because applying a differentiation matrix,
or local pseudopotential $V_{\eff}$ to different columns of $\Phi_K$
requires no communication. The inner products $\langle \cdot, \cdot
\rangle_{\mathcal{T}}$, however, are evaluated as vector inner
products (or a matrix-matrix multiplication when several integrals
are evaluated simultaneously). A block row partition is more
efficient for these types of operations. Collective communication is
required to sum up local products. To accommodate both data
distribution schemes in this step, we use the MPI\_Alltoallv call
within a global column processor group to convert $\Phi_K$ and
$\nabla \Phi_K$ from column partition to row partition.

%Such type of
%communication is all that is needed for operation type 1). No inter-element communication is needed. \CY{not sure why this is here}

%To see this let us take one term for example.
%\begin{equation}
%  \begin{split}
%    \average{\nabla \varphi_{K, j'}, \nabla \varphi_{K, j}}_{\mc{T}}
%    \equiv &
%    \int_{E_{K}} \nabla\varphi_{K, j'}(x) \nabla\varphi_{K, j}(x) \ud x\\
%    \approx&
%    \sum_{p=1}^{N_{g}^{\text{LGL}}} \varphi_{K, j'}(x)
%  \end{split}
%  \label{eqn:integralgradient}
%\end{equation}

%For inner products performed in
%Step 4), it is more efficient to use a row partition of $\Phi_K$  as
%shown in Figure~\ref{fig:DGDFT2}(b). We again use MPI\_Alltoallv to
%redistribute $\Phi_K$ when different types of operations need to be
%performed.

%To evaluate $\nabla \Phi_K$ on the LGL grid, we use a
%differentiation matrix along the $x,y,z$ directions, respectively.
%\CY{may need a bit more detail}

More sophisticated data communication is needed for step 2. This is
because the projection vector of a nonlocal pseudopotential may be
distributed among several elements (up to 8, see
Fig.~\ref{fig:HDG2D} (b)). If the nonzero support of the projection
operator is completely within one element $E_{K}$, this projection
operator only contributes to one diagonal block $H^{\DG}_{K;K}$.
Otherwise it contributes also to some off-diagonal matrix blocks
$H^{\DG}_{K';K}$ for all $E_{K'}$'s that contain a distributed
portion of the projection vector. Unlike the local pseudopotential
$V_{\eff}$, the nonlocal pseudopotential does not need to be updated
during the SCF iteration. Therefore, it is efficient for processors
associated with the element $E_{K}$ to own a distributed portion of
the projection vector $b_{I,\ell}(x)$ on the LGL grid constructed on
$E_{K}$, if $b_{I,\ell}(x)$ does not vanish on $E_{K}$. This allows
the inner product $\average{b_{I,\ell}, \varphi_{K, j}}_{\mc{T}}$ to
be entirely evaluated on processors associated with $E_{K}$ without
further inter-element communication. The computation of the matrix
element $\left(\sum_{I,\ell} \gamma_{I,\ell} \average{\varphi_{K',
j'}, b_{I,\ell}}_{\mc{T}} \average{b_{I,\ell}, \varphi_{K,
j}}_{\mc{T}} \right)$ however, requires data communication between
processors associated with $E_{K}$ and $E_{K'}$, on which
$b_{I,\ell}$ does not vanish. This is done by communicating the
inner products of the form $\average{b_{I,\ell}, \varphi_{K,
j}}_{\mc{T}}$ among such neighboring elements. Since the size of
such inner products is independent of the LGL grid size, the
communication volume of this step is relatively low. For
inter-element parallelization, we use asynchronous data
communication routines with MPI\_Isend and MPI\_Irecv for efficient
data communication.

The boundary integrals that appear in step 3 also require
inter-element communication. However, the inter-element data
communication only occurs among two neighboring elements
$E_{K},E_{K'}$ if the two elements share a common surface $S$.  It
should be noted that the computation of average and jump operators
only requires values of functions on the surface $S$, and therefore
only the function values of $\Phi_{K}$ and $\Phi_{K'}$ together with
their gradients $\nabla \Phi_{K}$ and $\nabla\Phi_{K'}$ restricted
to $S$ need to be computed.  These calculations require much lower
communication volume compared to those required in volume integrals.
We also remark that the communication performed in this step can be
overlapped with computation to further reduce the cost of
communication. In particular, we first launch the communication
needed to carry out step 3 before starting to perform the
computational tasks in step 1, which does not require inter-element
data communication.

Numerical results indicate that the overall procedure for
constructing the DG Hamiltonian matrix is highly efficient. For
instance, for the phosphorene P$_{3500}$ system, the total wall
clock time used to construct the DG Hamiltonian matrix is only 1.11
and 0.84 sec when the construction is carried out on 12,800 and
25,600 processors, respectively.

\textbf{Computation of the electron density}

As discussed in section~\ref{sec:dgintro}, the electron density can
be assembled from the eigenvector coefficients $\{c_{i;K,j}\}$, or
from the diagonal matrix blocks of the density matrix $P_{K;K}$
directly. Both options are supported in DGDFT. When the eigenvector
option is used, we diagonalize the DG Hamiltonian matrix (referred
to as the "DIAG" method) by using the ScaLAPACK software
package.\cite{ScaLAPACK} In order to use ScaLAPACK, we need to
convert the block row partition of the DG Hamiltonian matrix (see
Fig.~\ref{fig:HDG2D}) to the 2D block cyclic data distribution
scheme required by ScaLAPACK. The eigenvectors returned from
ScaLAPACK, which are stored in the 2D block cyclic format, are
redistributed according to the block row partition used to
distribute $H^{DG}$. We developed a routine to perform these
conversions. Such conversion is \textit{the only} operation in DGDFT
that involves MPI communication among, in principle, all the
available processors. All other MPI communication is performed
either within the global row processor group or the global column
processor group. We note that a similar conversion procedure also
can be found in other electronic structure software packages such as
SIESTA\cite{JPCM_14_2745_2002_SIESTA} and
CP2K\cite{JCTC_8_3565_2012} when ScaLAPACK is used.

Once the eigenvectors are redistributed by elements, the electron
density can be evaluated locally on each element. The global
electron density is imply the collection of the density defined on
each local element. Such global electron density is never collected
to be stored a single processor, but is distributed across
processors in a global row processor group.

It should be noted that for large systems, all ScaLAPACK
diagonalization routines (such as the divided and conquer routine
PDSYEVD currently used in DGDFT) have limited parallel scalability.
They often cannot make efficient use of the maximum number of
processors (which can be $10,000 \sim 100,000$ or more) that can be
used by other parts of the DGDFT calculation. Therefore, we often
need to restrict the ScaLAPACK calculation to a subset of processors
to avoid getting sub-optimal performance.

\subsection{Pole expansion and selected inversion method}\label{sec:pexsi}

When the electron density is computed from the expression given in
\eqref{eqn:newdensity}, we use the recently developed pole expansion
and selected inversion (PEXSI) method\cite{LinLuYingEtAl2009,
JPCM_25_295501_2013_PEXSI, JPCM_26_305503_2014_PEXSI} to compute the
diagonal blocks of the density matrix. This technique avoids the
diagonalization procedure which has an $\mathcal{O}(N^3)$
complexity. It is accurate for both insulating and metallic systems.
Furthermore, the computational complexity of the PEXSI method is
only $\mathcal{O}(N)$ for 1D systems, $\Or(N^{1.5})$ for 2D systems,
and $\mathcal{O}(N^2)$ for 3D systems. Therefore, the PEXSI method
is particularly well suited for studying electronic structures of
larges scale low-dimensional (1D and 2D)
systems.\cite{JCP_141_214704_2014_GNFs, PCCP_2015_ACPNRs}

The PEXSI method is based on approximating the density matrix
by a linear combination of Green's functions, i.e.,
\begin{equation}
  P \approx \sum_{l=1}^{Q} \mathrm{Im} \left[\omega_{l} (H^{\DG}-z_{l}I)^{-1} \right],
  \label{eqn:poleexpansion}
\end{equation}
where the integration weights $\omega_{l}$ and shifts $z_{l}$ are
chosen carefully so that the number of expansion terms $Q$ is
proportional to $\log (\beta\Delta E)$, where $\beta$ is the inverse
temperature and $\Delta E$ is the spectrum width of $H^{\DG}$. In
practical calculations, we observe that it is often sufficient to
choose $\Delta E$ to be much smaller than the true spectrum width of
$H^{\DG}$, thanks to the exponential decay of the Fermi-Dirac function
above the Fermi energy. In most cases, it is sufficient to choose
$Q=40\sim 80$. If we only need the diagonal blocks of the density
matrix $P$, we do not need to compute the entire inverse of
$H^{\DG}-z_{l}I$. Only the diagonal blocks of $(H^{\DG}-z_{l}I)^{-1}$
need to be computed, and these diagonal blocks can be computed
efficiently by using the selected inversion method.\cite{LinLuYingEtAl2009}
The use of selected inversion leads to favorable complexity of the PEXSI method.

The PEXSI method can scale to $10,000$ to $100,000$ processors. This
has recently been demonstrated in the massively parallel
SIESTA-PEXSI method\cite{JPCM_26_305503_2014_PEXSI,
JCP_141_214704_2014_GNFs}. SIESTA-PEXSI uses local atomic orbitals
to discretize the Kohn-Sham Hamiltonian. Because DGDFT is designed
to take advantage of massively parallel computers, the high
scalability of PEXSI, in addition to its lower asymptotic
complexity, makes it a more attractive option compared to the
diagonalization option. When the PEXSI option is used, DGDFT can
often scale to $10,000 \sim 100,000$ processors, and be used to
solve the electronic structure problems with more than $10,000$
atoms efficiently.\cite{PCCP_2015_ACPNRs}

In order to use PEXSI in DGDFT, the DG Hamiltonian matrix needs to
be redistributed from a block row distribution format to a
distributed compressed sparse column format (DCSC). When the DCSC
format is used to distribute a sparse Hamiltonian matrix among $M$
processors, each processor holds roughly $\lfloor n/M\rfloor$
columns stored in a compressed sparse column (CSC) format, where $n$
is the dimension of the matrix. The diagonal blocks of the density
matrix returned from PEXSI, which are stored in DCSC format, are
converted back to the block row partition format used to represent
the electron density.

Because the selected inversions of the shifted DG Hamiltonians
$(H^{\DG}-z_{l}I)$ are independent of each other for different poles
$z_{l}$, we can carry out these selected inversions among different
global row processor groups. Therefore, unlike the data
redistribution scheme used for ScaLAPACK. As a result, the procedure
for redistributing $H^{DG}$ required by PEXSI uses asynchronous
communication only within a global row processor group.

\begin{figure}[htbp]
\begin{center}
\includegraphics[width=0.5\textwidth]{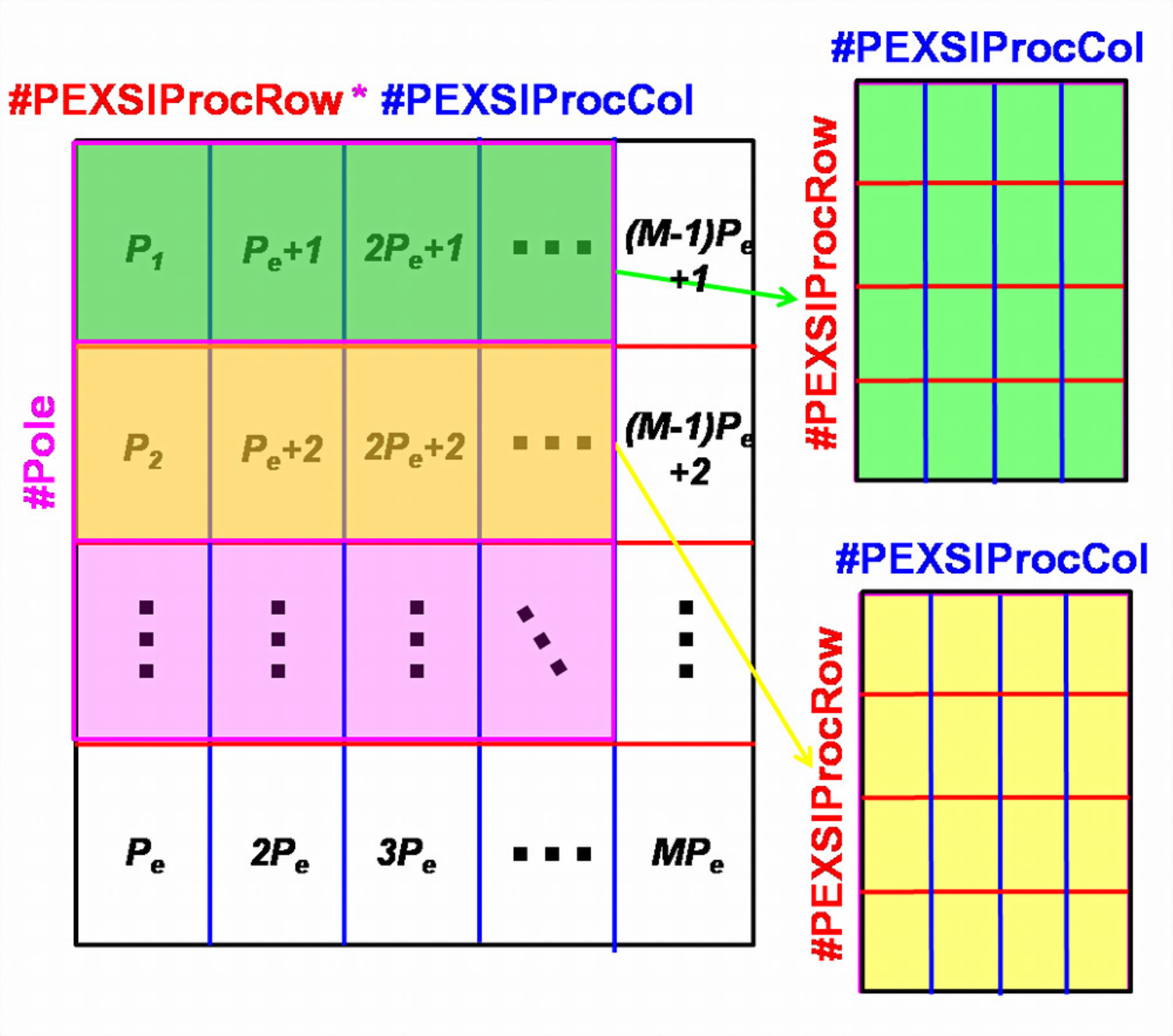}
\end{center}
\caption{(Color online)
The processor grid used in the PEXSI method. The processors used
in PEXSI is usually a subset of all available processors.
\#Pole represents the number of poles used in the
PEXSI method. The inset shows that each 1D global row processor group is
further partitioned into a 2D processor group used for each selected
inversion.
%Distribution of two levels of
%parallelization of the PEXSI method in the DGDFT methodology. \#Pole
%represents the number of poles used in the PEXSI method. The inset
%shows that each 1D global row processor group is further partitioned
%into a 2D processor group inside PEXSI.
} \label{fig:PEXSI}
\end{figure}

Because each parallel selected inversion requires processors to be
arranged logically into a 2D grid of size
(\#PEXSIProcRow)$\times$(\#PEXSIProcCol), the processors within each
global row processor group is reconfigured in PEXSI as shown in
Fig.~\ref{fig:PEXSI}. Note that restricting the selected inversion
at each pole to processors belonging to a global processor row group
limits the maximum number of processors that can be used for each
selected inversion to the number of elements $M$. However, for
systems of large size, $M$ can be easily $1000$ or more. Therefore
such a configuration does not severely limit the number of
processors that can be effectively used for selected inversion. If
the number of processors ($P_{e}$) within each global column
processors group is larger than the number of poles $Q$, all
selected inversions required in the pole
expansion~\eqref{eqn:poleexpansion} can be carried out
simultaneously. The number of processors that can effectively be
used in PEXSI is $MQ$. If $Q > P_{e}$, we have to process at most
$P_{e}$ poles at a time, and repeat the process until selected
blocks of $(H-z_l I)^{-1}$ have been computed for all $z_l$'s. In
this case and when all $M$ processors in a global processor row
group are used for selected inversion, all $M P_{e}$ processors can
be effectively used by PEXSI.

\section{Computational results}
\label{sec:results}

In this section, we report the performance and accuracy of DGDFT
when it is applied to 2D phosphorene systems of different sizes.

Phosphorene, a new two dimensional (2D) elemental
monolayer,\cite{ACSNano_8_4033_2014, NatureCommun_5_4475_2014,
NatureNanotech_9_372_2014, JPCL_5_1289_2014, JMCC_3_4756_2015} has
received considerable amount of interest recently after it has been
experimentally isolated through mechanical exfoliation from bulk
black phosphorus. Phosphorene exhibits some remarkable electronic
properties superior to graphene, a well known elemental
sp$^2$-hybridized carbon monolayer.\cite{Scinece_306_666_2004,
NatureMater_6_183_2007, RMP_81_109_2009} For example, phosphorene is
a direct semiconductor with a high hole
mobility.\cite{ACSNano_8_4033_2014} It has the drain current
modulation up to 10$^5$ in
nanoelectronics.\cite{NatureCommun_5_4475_2014} These remarkable
properties have already been used for wide applications in field
effect transistors\cite{NatureNanotech_9_372_2014} and thin-film
solar cells.\cite{JPCL_5_1289_2014} Furthermore, up to now,
phosphorene is the only stable elemental 2D material which can be
mechanically exfoliated in experiments\cite{ACSNano_8_4033_2014}
besides graphene. Therefore, it can potentially be used as an
alternative to graphene in the future and lead to faster
semiconductor electronics.

Fig.~\ref{fig:Structure} shows the atomic configuration of 2D
phosphorene monolayer in a 5 $\times$ 7 supercell (P$_{140}$). Other
phosphorene models in very large supercells involving thousands or
tens of thousands of atoms, such as the 25 $\times$ 35 (P$_{3500}$)
and 50 $\times$ 70 (P$_{14000}$) supercells, which we use as test
problems in this work, are not shown here. The vacuum space in the X
and Y directions is about 10 {\AA} to separate the interactions
between neighboring slabs in phosphorene.

All calculations are performed on the Edison platform available at
the National Energy Research Scientific Computing (NERSC) center.
Edison has 5462 Cray XC30 nodes. Each node has 24 cores partitioned
among two Intel Ivy Bridge processors. Each 12-core processor runs
at 2.4GHz, and has 64 GB of memory per node. The maximum number of
available cores is 131,088 on Edison. In all calculations, we
utilize all $24$ cores on a computational node.

%\LL{In the report below we need to change all
%``processors'' into ``cores'' to be consistent with the description
%here. This issue has recently been picked out by a referee.}
%There are 5462 computer nodes and 24 processors in each node on
%Edison with 64 GB of memory per node. The maximum number of
%available processors is 131,088 on Edison.
%\CY{give some more machine specs}

\subsection{Computational accuracy}

We use the conventional plane wave software package
ABINIT\cite{CPC_180_2580_2009_ABINIT} as a reference to check the
accuracy of results from DGDFT. The same exchange-correlation
functional of the local density approximation of Goedecker, Teter,
Hutter (LDA-Teter93)\cite{PRB_54_1703_1996_LDA} and the
Hartwigsen-Goedecker-Hutter (HGH) norm-conserving
pseudopotential\cite{PRB_58_3641_1998_HGH} are adopted in both
ABINIT and DGDFT software packages.

We first check the accuracy of the total energy and the atomic force
of the DGDFT method by using P$_{140}$ shown in
Fig.~\ref{fig:Structure} as an example. To simplify our discussion,
we define the total energy error per atom $\Delta$E (Hartree/atom)
and maximum atomic force error $\Delta$F (Hartree/Bohr) as
\[
\Delta{E}=(E^{\text{DGDFT}}-E^{\text{ABINIT}})/N_{A},
\] and
\[
\Delta{F}=\max_I|F_I^{\text{DGDFT}}-F_I^{\text{ABINIT}}|.
\]
respectively, where $N_{A}$ is the total number of atoms.
%and $I$ correspond to the total number of
%atoms and an atom index,
$E^{\text{DGDFT}}$ and $E^{\text{ABINIT}}$ represent the total
energy computed by DGDFT and ABINIT respectively, and
$F_I^{\text{DGDFT}}$ and $F_I^{\text{ABINIT}}$ represent the
Hellmann-Feynman force on the $I$-th phosphorus atom in P$_{140}$
computed by DGDFT and ABINIT, respectively. The ABINIT results are
obtained by setting the energy cutoff to 200 Hartree for the
wavefunction to ensure full convergence. The kinetic energy cutoff
(denoted by Ecut) in the DGDFT method is used to defined the grid
size for computing the ALBs as is in standard Kohn-Sham DFT
calculations using planewave basis sets. Ecut is also directly
related to the Legendre-Gauss-Lobatto (LGL) integration grid defined
on each element and used to perform numerical integration as needed
to construct the DG Hamiltonian matrix. The number of LGL grids per
direction is set to be twice the number of grid points calculated
using Eq.~\eqref{eqn:Ngecut} with the same Ecut.

Table~\ref{Accuracy} shows that the total energy and atomic forces
produced by the DGDFT method are highly accurate compared to the ABINIT
results. In particular, the total energy error $\Delta E$ can be as
small as $3.39 \times 10^{-6}$ Hartree/atom if the DIAG method is used
and $8.12 \times 10^{-5}$ Hartree/atom if the PEXSI method is used
respectively. Here $50$ poles are used and the accuracy of PEXSI can be
further improved by increasing the number of poles.%\LL{please check}
The maximum error of the atomic force can be as small as $1.06
\times 10^{-4}$ Hartree/Bohr when DIAG is used and $1.06 \times
10^{-4}$ Hartree/Bohr when PEXSI is used. These results are obtained
when only a relatively small number of ALB functions per atom are
used to construct the global DG Hamiltonian. The energy cutoff for
constructing the ALBs is set to 200 Hartree in this case.  Note that
the accuracy of total energy and atomic force in DGDFT depends on
both the energy cutoff for local wavefunctions defined on an
extended element and the number of ALB functions. We can see from
Table~\ref{Accuracy} that the accuracy of the total energy and
atomic forces both improve as the energy cutoff and the number of
ALB functions increases.  We also find that the use of the
Hellmann-Feynman force can result in accurate force calculation,
despite that the contribution from the Pulay force\cite{Pulay1980}
is not included.
\begin{table}
\caption{The accuracy of DGDFT in terms of the total energy error
per atom $\Delta$$E$ (Hartree/atom) and the maximum atomic force
error $\Delta$$F$ (Hartree/Bohr) in the DIAG and PEXSI methods with
different kinetic energy cutoff Ecut (Hartree), and number of ALB
functions per atom, compared with converged ABINIT calculations.
\#ALB means the number of ALB functions per atom. } \label{Accuracy}
\begin{tabular}{cccccc} \\ \hline \hline
\multicolumn{2}{c}{DGDFT P$_{140}$}  & \multicolumn{2}{c}{DIAG}  & \multicolumn{2}{c}{PEXSI} \ \\
Ecut &  \#ALB  & $\Delta$$E$ & $\Delta$$F$ & $\Delta$$E$ & $\Delta$$F$  \ \\
\hline
20   &  91.43   & 5.22E-04  &  4.03E-03  &  3.22E-04  &  4.03E-03  \ \\
40   &  18.28   & 4.51E-02  &  5.97E-02  &  4.57E-02  &  5.97E-02  \ \\
40   &  27.42   & 6.67E-04  &  2.51E-03  &  6.85E-04  &  2.52E-03  \ \\
40   &  36.57   & 1.34E-04  &  6.16E-04  &  1.59E-04  &  6.18E-04  \ \\
40   &  45.71   & 7.00E-05  &  4.00E-04  &  6.44E-05  &  5.23E-04  \ \\
40   &  91.43   & -4.32E-07 &  5.93E-04  &  1.34E-04  &  5.93E-04  \ \\
100  &  91.43   & 1.59E-05  &  1.97E-04  &  8.04E-05  &  1.97E-04  \ \\
200  &  91.43   & 3.39E-06  &  1.06E-04  &  8.12E-05  &  1.06E-04  \ \\
\hline \hline
\end{tabular}
\end{table}

In the following parallel efficiency tests, we set the energy cutoff
to 40 Hartree for ALBs and 36.57 ALB functions per atom (80
ALB functions per element), which achieves good compromise between
accuracy and computational efficiency. For this particular choice of
the energy cutoff and the number of ALB functions, we are able to keep the
total energy error under 1 $\times$ 10$^{-4}$ Hartree/atom and
atomic force error under 1 $\times$ 10$^{-3}$ Hartree/Bohr for 2D
phosphorene systems.
%\LL{Just to confirm: Checked that such
%accuracy is satisfied for larger systems?}

\subsection{Parallel efficiency}

To illustrate the parallel scalability of the DGDFT method, we
demonstrate the performance of three main steps in each SCF
iteration as shown in Fig.~\ref{fig:Flowchart}: (a) the generation
of ALB functions, (b) the construction of DG Hamiltonian matrix and
(c) the evaluation of the approximate charge density, energy and
atomic forces by either diagonalizing the DG Hamiltonian (DIAG) or
by using the PEXSI technique. Note that there are some additional
steps such as the computation of energy, charge mixing or potential
mixing, and intermediate data communication etc. The cost of these
extra steps is included in the total wall clock time.

\begin{figure}[htbp]
\begin{center}
\includegraphics[width=0.5\textwidth]{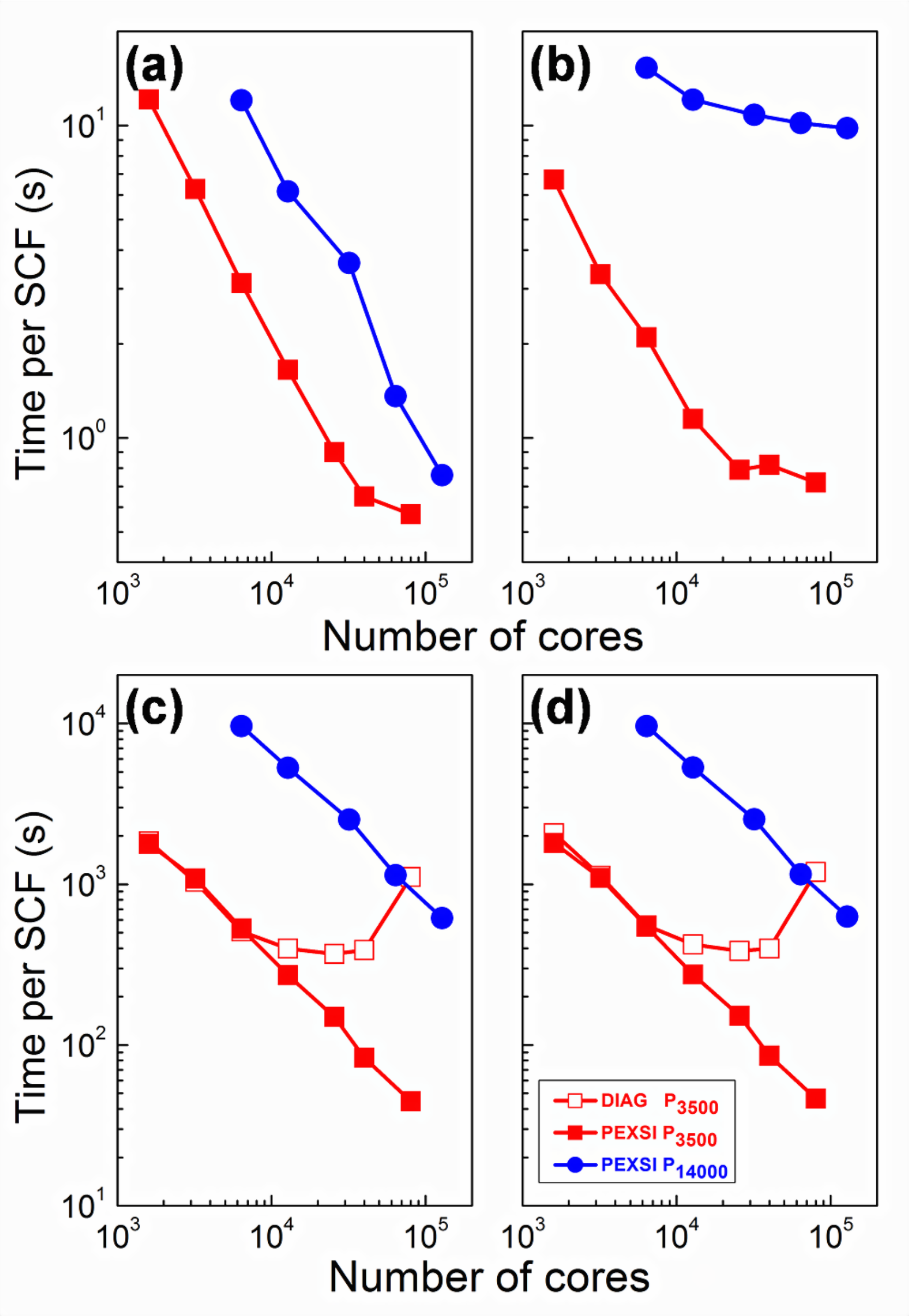}
\end{center}
\caption{(Color online) The wall clock time with respect
to the number of cores used for the computation for 2D
phosphorene monolayer systems of different sizes (P$_{3500}$ and
P$_{14000}$). Strong scaling of (a) the generation of ALB functions
step, (b) the DG Hamiltonian matrix construction step, (c) the
evaluation of the approximate charge density, energy and forces from
the constructed DG Hamiltonian matrix, (d) the overall computational
time.  The reported wall clock time is for one SCF iteration. The timing
and scaling shown in (c) and (d) depend on whether DIAG (hollow
markers) or PEXSI (solid markers) is used to evaluate physical
quantities such as charge density, energy and forces. }
\label{fig:Time}
\end{figure}

Fig.~\ref{fig:Time} shows the strong parallel scaling of these three
individual steps, as well as the overall DGDFT method, for two
large-scale 2D phosphorene monolayers (P$_{3500}$ and P$_{14000}$)
in terms of the wall clock time per SCF step. For P$_{3500}$, we
tested the performance using both the DIAG and PEXSI methods for the
evaluation of charge density during SCF. But for P$_{14000}$, we
only use the PEXSI method, since the DIAG method is too expensive
for systems of such size (the dimension of the matrix is $512,000$).

The wall clock time of the first two steps are independent of
whether PEXSI or DIAG is used. Fig.~\ref{fig:Time}(a) and (b) show
that they both scale well with respect to the number of cores used
in the computation for all test problems we used. The exception is
the construction of the DG Hamiltonian which does not scale beyond
$10^{4}$ processors for P$_{14000}$. We find that the only routine
does not scale well is the inter-element communication of the
boundary values of $\Phi_{K}$ and $\nabla \Phi_{K}$, which is
currently implemented via asynchronous communication. Since the
volume of the asynchronous communication is proportional to the
system size, for a large system the communication volume may exceed
the size of the MPI buffer which leads to sub-optimal performance of
the asynchronous data communication. Nonetheless, the cost of the
generation of ALBs and the construction of the DG matrix is much
less compared to that for computing the electron density from the DG
Hamiltonian.

%\LL{Which part in
%the construction of the DG matrix does not scale well for the 14000
%atom case?}
%The total wall clock time required to perform each one
%these steps is reduced to a few seconds even for P$_{14000}$.

Fig.~\ref{fig:Time}(c) and (d) show that the evaluation of the
approximate charge density using the DG Hamiltonian matrix dominates
the total wall clock time per SCF iteration in the DGDFT methodology
for systems of large sizes. For large-scale 2D phosphorene systems
P$_{3500}$ and P$_{14800}$, the PEXSI method can effectively reduce
the wall clock time compared to the DIAG method in the DGDFT
methodology. Furthermore, using the DIAG method with
ScaLAPACK,\cite{ScaLAPACK} appears to limit the strong parallel
scalability to at most 10,000 cores on the Edison. Increasing the
cores beyond that can lead to an increase in wall clock time. In
contrast, the PEXSI method exhibits highly scalable performance. It
can make efficient use of about 100,000 cores on Edison for
P$_{3500}$ and P$_{14000}$.

\begin{figure}[htbp]
\begin{center}
\includegraphics[width=0.5\textwidth]{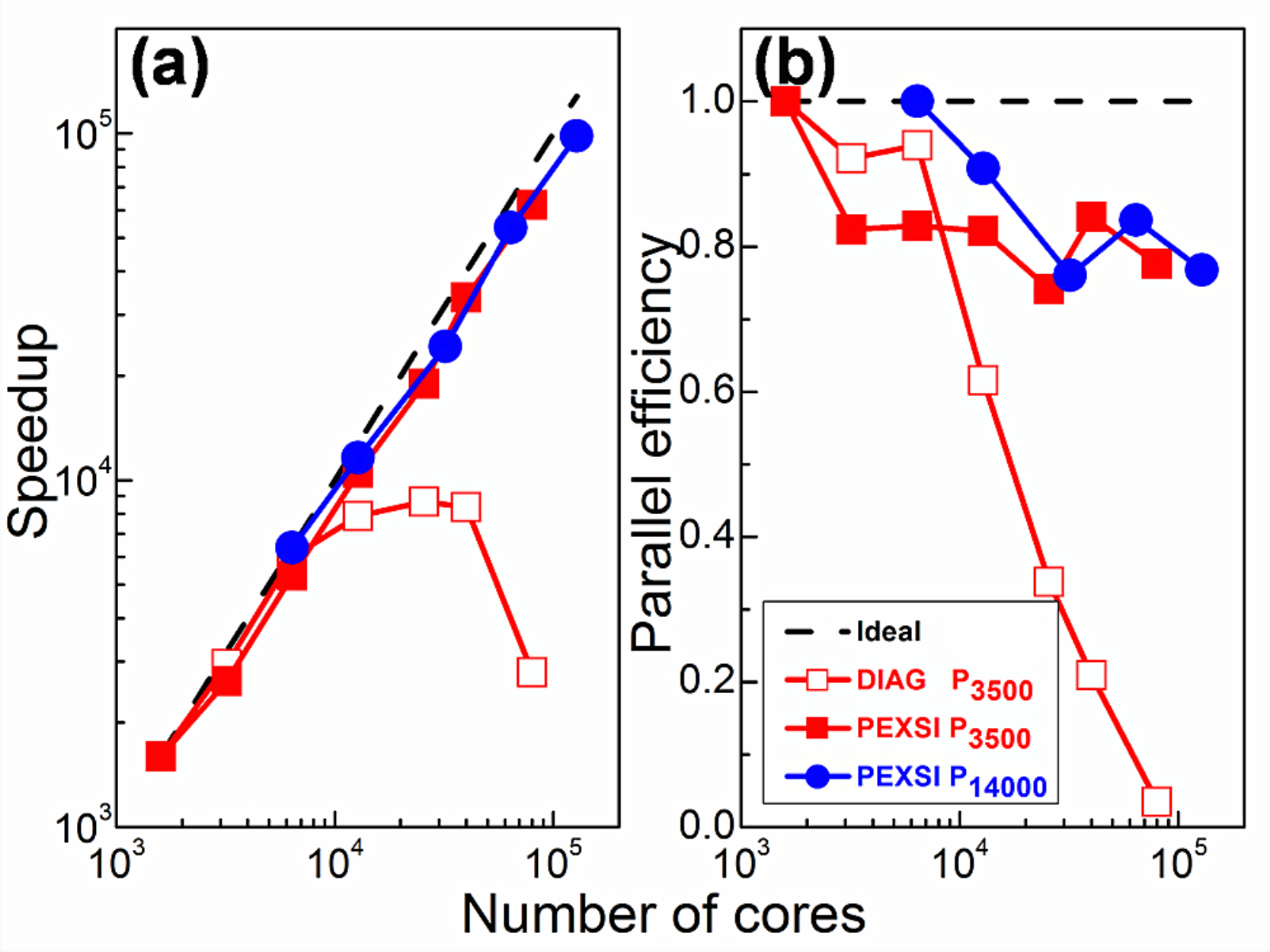}
\end{center}
\caption{(Color online) The (a) speedup and (b) parallel efficiency
with respect to the number of cores used for the computation for 2D
phosphorene monolayer systems of different sizes (P$_{3500}$ and
P$_{14000}$).} \label{fig:Speedup}
\end{figure}

Fig.~\ref{fig:Speedup} shows the speedup and parallel efficiency
with respect to the number of cores used for the computation for 2D
phosphorene P$_{3500}$ and P$_{14000}$. For the P$_{3500}$ system,
both the DIAG and PEXSI methods in DGDFT can keep highly parallel
efficiency (90\% for DIAG and 80\% for PEXSI) with less than 10,000
cores. But for the DIAG method, further increase of the number of
cores will lead to rapid loss of parallel efficiency. On the
contrary, the PEXSI method is highly scalable, and its parallel
efficiency is about 80\% even when the number of cores increases
beyond 80,000 for the P$_{3500}$ system. For the large-scale
P$_{14000}$ system, we only use the PEXSI method in DGDFT, and we
find that the parallel efficiency of the DGDFT-PEXSI method around
80\% when 128,000 cores are used on Edison (Edison has $131,088$
cores in total).

\section{Conclusions}\label{sec:conclusion}

We described a massively parallel implementation of the DGDFT
(Discontinuous Galerkin Density Functional Theory) methodology that
can be used to perform large-scale Kohn-Sham density functional
theory (DFT) calculations efficiently. We demonstrated the accuracy
and efficiency of our parallel implementation. In particular, we
showed that DGDFT can achieve accuracy comparable to that produced
by a conventional planewave based calculation with far fewer number
of basis functions. We also showed that DGDFT can efficiently use
128,000 computational cores to solve a problem with over $10,000$
atoms. The high parallel efficiency results from a two-level
parallelization schemes that make use of several different types of
data distribution, task scheduling and data communication schemes.
It also benefits from the use of the PEXSI method to compute
electron density, energy and atomic forces. The PEXSI method has a
favorable computational complexity and is also amenable to a
two-level parallelization scheme that enables it to achieve high
parallel efficiency.

\section{Acknowledgement}

This work is partially supported by the Scientific Discovery through
Advanced Computing (SciDAC) Program funded by U.S. Department of
Energy, Office of Science, Advanced Scientific Computing Research
and Basic Energy Sciences (W. H., L. L. and C. Y.), and by the
Center for Applied Mathematics for Energy Research Applications
(CAMERA), which is a partnership between Basic Energy Sciences and
Advanced Scientific Computing Research at the U.S Department of
Energy (L. L. and C. Y.). We thank the National Energy Research
Scientific Computing (NERSC) center for the computational resources.

\footnotesize{
\bibliography{achemso}
}

\end{document}